\documentclass[aps,pra,showpacs]{revtex4}
\usepackage{amsmath}
\usepackage{graphicx}
\usepackage{dcolumn}
\usepackage{longtable}

\newcommand{\be}{\begin{equation}}
\newcommand{\ee}{\end{equation}}
\newcommand{\bearray}{\begin{eqnarray}}
\newcommand{\eearray}{\end{eqnarray}}
\newcommand{\bse}{\begin{subequations}}
\newcommand{\ese}{\end{subequations}}

\newcommand{\fourth}{\frac{1}{4}}

\begin{document}

\title{Positronium energy levels at order $m \alpha^7$: light-by-light scattering in the two-photon-annihilation channel}

\author{Gregory S. Adkins}
\email[]{gadkins@fandm.edu}
\author{Christian Parsons}
\author{M. D. Salinger}
\author{Ruihan Wang}
\affiliation{Franklin \& Marshall College, Lancaster, Pennsylvania 17604}
\author{Richard N. Fell}
\affiliation{Brandeis University, Waltham, Massachusetts 01742}

\date{\today}

\begin{abstract}
Recent and ongoing experimental work on the positronium spectrum motivates new efforts to calculate positronium energy levels at the level of three loop corrections.  We have obtained results for one set of such corrections involving light-by-light scattering of the photons produced in a two-photon virtual annihilation process.  Our result is an energy shift $1.58377(8) m \alpha^7/\pi^3$ for the $n=1$ singlet state, correcting the ground state hyperfine splitting by $-6.95 kHz$.  We also obtained a new and more precise result for the light-by-light scattering correction to the real decay of parapositronium into two photons.
\end{abstract}


\maketitle


\section{Introduction}
\label{introduction}

Positronium has interesting and unique features that make it a crucial test case in our quest to understand the application of quantum field theory to physical systems.  In comparison to hydrogen, positronium is far simpler in that no constituent structure effects enter the picture.  For hydrogen, the leading uncertainty in, say, the ground state hyperfine splitting (hfs) is due to proton structure \cite{Ramsey93,Eides07}.  In fact, such structure effects have been implicated recently in a $7 \sigma$ difference between the values of the proton charge radius as determined in normal hydrogen and muonic hydrogen \cite{Antognini13,Jentschura11}.  High precision studies in positronium and also muonium augment our understanding of the application of quantum electrodynamics (QED) to bound systems without this added complication of structure.  On the other hand, positronium physics involves some significant complications not present in hydrogen, specifically the lack of a mass-scale expansion parameter and the presence of virtual annihilation effects.  Hydrogen physics is significantly simplified in the no-recoil limit and expansion in the electron to proton mass ratio is a useful tool for performing high-precision calculations.  Due to the equality of the electron and positron masses there is no such simplification in positronium--recoil effects are as large as they could be.  Furthermore, because of this equality of masses there is no suppression of the effect of the magnetic moment of the positively charged particle in positronium as there is in hydrogen.  Thus the positronium hfs is not suppressed relative to the fine structure--both contribute on an equal footing at order $m \alpha^4$.  Virtual annihilation effects are another characteristic feature of positronium.  These effects can be very important--for example, for the ground state hfs, virtual annihilation supplies $3/7$ of the leading order effect.  The study of binding in positronium, muonium, hydrogen, and a number of additional exotic atoms \cite{Eides07,Karshenboim04,Karshenboim05} complements other non-binding applications of QED such as the electron and muon magnetic moments \cite{Hanneke11,Kinoshita14,Bennett06,Aoyama12} to advance our understanding of quantum field theory in general and lead to improved values for the fundamental constants.  Confirmed discrepancies between predicted and measured values could be a signal for new physics beyond the standard model.

Positronium was first produced in 1951 \cite{Deutsch51}, and since then increasingly precise measurements have been made of the ground state hfs, orthopositronium (spin-triplet) and parapositronium (spin-singlet) decay rates and branching ratios, $n=2$ fine structure, and the $2S-1S$ interval.  (Reviews of this progress can be found in Refs~\cite{Berko80,Rich81,Mills90,Rich90,Dvoeglazov93,Dobroliubov93,Karshenboim04,Rubbia04,Gninenko06}.)  The most precise hfs measurements date from the early 80s \cite{Mills75,Mills83,Ritter84}:
\bse
\bearray
\Delta E(\rm{Brandeis}) &=& 203 \, 387.5(1.6) MHz , \cr
\Delta E(\rm{Yale}) &=& 203 \, 389.10(74) MHz .
\eearray
\ese
There has been a significant amount of recent activity directed towards new measurements of positronium transition energies \cite{Fan96,Crivelli11,Sasaki11,Ishida12,Yamazaki12, Namba12, Cassidy12} and a new result for the hfs has been reported \cite{Ishida14}
\be
\Delta E(\rm{Tokyo}) = 203 \, 394.2 (1.6)_{\rm{stat}} (1.3)_{\rm{sys}} MHz .
\ee

The corresponding theoretical work on the hfs was reviewed in \cite{Adkins14}.  The present theoretical result can be expressed as
\be
\Delta E = m \alpha^4 \Bigl \{ C_0 + C_1 \frac{\alpha}{\pi} + C_{21} \alpha^2 L + C_{20} \left ( \frac{\alpha}{\pi} \right )^2 + C_{32} \frac{\alpha^3}{\pi} L^2 + C_{31} \frac{\alpha^3}{\pi} L + C_{30} \left ( \frac{\alpha}{\pi} \right )^3 + \cdots \Big \}
\ee
where $L = \ln (1/\alpha)$.  All coefficients in this expression are known (analytically) except for $C_{30}$ representing the three-loop non-logarithmic terms.  Only a part of $C_{30}$ is presently known.  Various contributions to $C_{30}$ involve virtual annihilation (into one, two, three, or four photons) or no virtual annihilation (radiative, radiative-recoil, and recoil corrections).  Corrections can also be distinguished by the significant scales that enter the dynamics: hard (the scale of the electron mass $m$), soft (that of the bound-state momentum $m \alpha$), and ultrasoft (the scale of bound-state energies $m \alpha^2$).  Despite the fact that the naive magnitude of three-loop corrections is only $m \alpha^4 (\alpha/\pi)^3 = 4.39 kHz$, large contributions to $C_{30}$ have been found.  Marcu found a part of $C_{30}$ having value $\sim 109$ arising from ultrasoft contributions, leading to a significant energy correction $0.48 MHz$ \cite{Marcu11}.  A complete calculation of the one-photon-annihilation contribution (including all energy scales)  was recently reported by Baker {\it et al.} \cite{Baker14} to make a contribution $0.217(1) MHz$ to the hfs.  Additionally, a number of hard-scale contributions have been obtained by Adkins and Fell \cite{Adkins14} and by Eides and Shelyuto \cite{Eides14}, involving light-by-light scattering in the two-photon-exchange channel and radiative corrections to two-photon exchange.  The present work is a calculation of another contribution to $C_{30}$ coming from the two-photon-annihilation channel with an intermediate light-by-light scattering.  This light-by-light contribution is the only three-loop annihilation graph that can be cut in more than one way into virtual particles (pairs of photons) that, in the proper kinematic region, can be on the mass shell.

Contributions to positronium energy shifts that involve virtual annihilation to two or more photons produce complex energy corrections.  The real parts of these corrections contribute to energies in the usual way; imaginary parts give partial rates for decay into the corresponding channels according to
\be \Delta \Gamma = -2 \, \mathrm{Im} ( \Delta E ). \label{rate_from_ImE}\ee  
The presence of these imaginary parts is both a help and a hindrance.  They significantly complicate the analysis, as the real and imaginary parts of each contribution must be separated at some point in the analysis--certainly before performing numerical integration to obtain final results.  On the other hand, the corresponding decay rate contributions are known from prior work and supply useful checks on the new calculations.

The regions in phase space that give rise to imaginary parts are known from Cutkosky analysis \cite{Cutkosky60,Veltman94,Peskin95}: they are the regions for which the virtual photons can become real.  That is, the photons of momenta $k_i$ (for $2\le i \le n$ with $n\ge2$ must be real ($k_i^2=0$) with positive energy ($k_i^0 > 0$) and have total four-momentum equal to that of the positronium state.  (In the center-of-mass frame this last condition reads $k_1^0 + \cdots k_n^0 = E_{ps}$ where $E_{ps} = 2m-m \alpha^2/(4 n^2)+ O(m \alpha^4) \approx 2m$.)  The light-by-light graphs that are the subject of this paper possess two simultaneous cuts, as can be seen on any of the diagrams of Fig.~1.  They are unique among three-loop positronium energy corrections in this important aspect.  As mentioned above, the imaginary parts of these graphs are related to the corrections to two-photon decay of parapositronium due to light-by-light scattering in the final state.  These were computed previously as part of the calculation all two-loop corrections to the parapositronium decay rate. \cite{Czarnecki99,Adkins01}

In this paper we describe our calculation of the positronium energy correction due to light-by-light scattering in the two-photon-annihilation channel.  General aspects of the calculation are discussed in Sec.~\ref{calc}, after which the six light-by-light diagrams are divided into two classes--planar and crossed--that are evaluated separately in Sections \ref{planar} and \ref{crossed}.  In Sec.~\ref{results} we summarize our results.


\section{General aspects of the calculation}
\label{calc}

\begin{figure}
\includegraphics[width=6.5in]{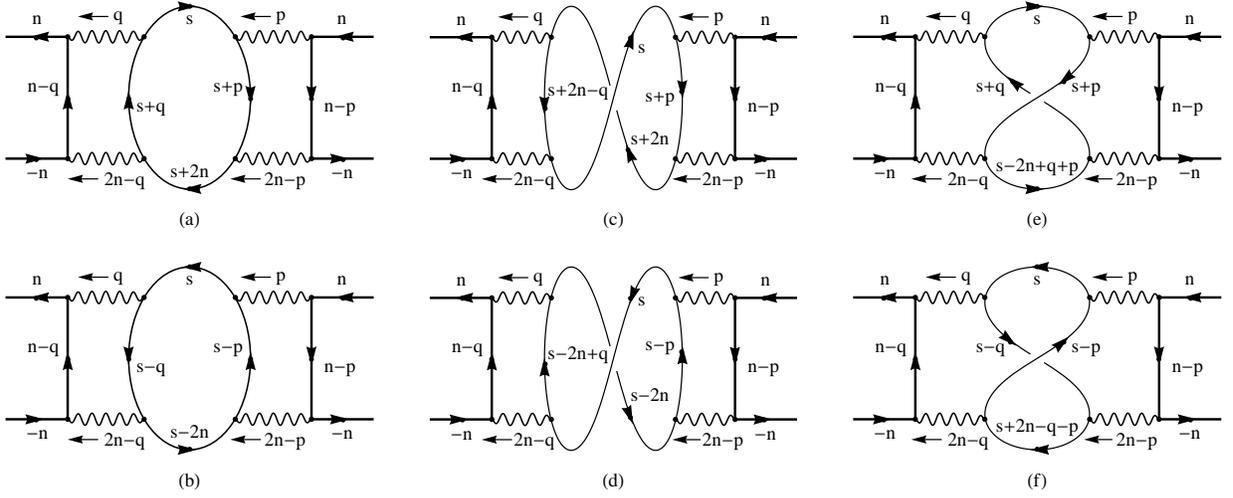}
\caption{\label{fig1} The six light-by-light scattering graphs in the two-photon-annihilation channel.  The six graphs represent the six possible ways that positronium states on the left and right can be connected by two-photon intermediate states with a light-by-light scattering process in between.  The graphs on the bottom row are identical to those on the top except that the light-by-light loop is traversed in the opposite direction.  The vector $n=(1,\vec 0\,)$ is used to represent half of the positronium energy-momentum in units of the electron mass $m$.  Momentum labels on fermion lines by convention flow in the direction of the arrow on the line.}
\end{figure}

The diagrams that contribute to the light-by-light scattering correction in the two-photon-annihilation channel are shown in Fig.~\ref{fig1}.  The positronium states on the left and right are represented by their constituent electrons and positrons.  The six diagrams represent the six ways that the closed electron loop can connect to four virtual photons.  Momenta are labeled so that these six diagrams are identical outside of the closed electron loop.  The momentum four-vector $n$ here represents half of the total energy-momentum of the positronium state (in units of the electron mass $m$), and use of the center-of-mass frame is assumed.  The energy level contribution from this set of diagrams is finite both in the infrared and ultraviolet.  The light-by-light contribution is ``hard''--that is, it involves virtual momenta of the order of the electron mass and not from the soft or ultrasoft scales.  Consequently, the energy shift at leading order vanishes unless the electron and positron can be in spatial contact, {\it i.e.} unless the orbital angular momentum $\ell$ is zero.

The light-by-light scattering process takes place through a two-photon intermediate state with charge conjugation quantum number $C=+1$, so the positronium states that are involved must also have $C=+1$ where $C=(-1)^{\ell+s}$ for positronium with total spin $s$.  That is, the affected states are the spin-singlet states with $s=0$.  The order of this three-loop contribution is $m \alpha^7$, with $\alpha^3$ coming from the square of the wave function at the origin and $\alpha^4$ from the eight powers of the electron charge $e$ explicitly present according to the Feynman rules.

The energy shift due to these light-by-light processes can be found in a number of ways.  We choose to use a Bethe-Salpeter based formalism because it leads very directly to an answer for a process of this sort that lacks dependence on the soft or ultrasoft energy scales.  The energy shift, in the formalism described in Ref.~\cite{Adkins99}, is an expectation value of an interaction operator between bound state wave functions, that is
\be \Delta E = i \bar{\Psi} \delta K \Psi \ee
where $\delta K$ represents the light-by-light scattering operator (including attached photons and the electrons that connect them) and $\bar{\Psi}$ and $\Psi$ are the positronium wave functions.  The most basic approximations for the wave functions are sufficient for our purposes here:
\be \Psi \rightarrow (2 \pi)^4 \delta^4 (\ell) \phi_0 \begin{pmatrix} 0 & \chi \\ 0 & 0 \end{pmatrix}, \quad \bar \Psi^T \rightarrow (2 \pi)^4 \delta^4 (\ell) \phi_0 \begin{pmatrix} 0 & 0 \\ \chi^\dagger & 0 \end{pmatrix} \ee
where $\ell$ is the relative momentum, $\phi_0=\sqrt{m^3 \alpha^3/(8 \pi n^3)}$ is the wave function at spatial contact for a state of principal quantum number $n$ and $\ell=0$, and $\chi$ is the Pauli $2 \times 2$ spin state.  We require only the form of $\chi$ for parapositronium (total spin 0): $\chi \rightarrow 1/\sqrt{2}$.  The explicit expression for the energy shift for, say, the diagram of Fig.~\ref{fig1}a is
\bearray
\Delta E_a &=& (-1) i \phi_0^2 \int \frac{d^4 p}{(2 \pi)^4} \frac{d^4 q}{(2 \pi)^4} \frac{d^4 s}{(2 \pi)^4}  \mathrm{tr} \Bigl [ (-i e \gamma^\alpha) \frac{i}{\gamma(P/2-q)-m} (-i e \gamma^\beta) \begin{pmatrix} 0 & 0 \\ \chi^\dagger & 0 \end{pmatrix} \Bigr ] \cr
&\times& \frac{-i}{p^2} \frac{-i}{(P-p)^2} \frac{-i}{q^2} \frac{-i}{(P-q)^2} 
 \mathrm{tr} \Bigl [ (-i e \gamma^\mu) \frac{i}{\gamma(P/2-p)-m} (-i e \gamma^\nu) \begin{pmatrix} 0 & \chi \\ 0 & 0 \end{pmatrix} \Bigr ] \cr
&\times&  (-1) \; \mathrm{tr} \Bigl [ (-i e \gamma_\mu) \frac{i}{\gamma(s+p)-m} (-i e \gamma_\nu) \frac{i}{\gamma s-m} (-i e \gamma_\alpha) \frac{i}{\gamma(s+q)-m} (-i e \gamma_\beta) \frac{i}{\gamma (s+P)-m} \Bigr ] .
\eearray
Here $P$ is the positronium energy-momentum vector.  In the center of mass frame we write this as $P = 2 m n$ where $n=(1,\vec 0 \,)$ is a timelike unit vector.  We include two factors of $-1$ coming from Fermi symmetry because this graph contains two annihilations into virtual photons.  We can simplify this expression by rationalizing fermion propagation factors and using projection operators to write the spin matrices in terms of gamma matrices (in the Dirac representation) \cite{Itzykson80}:
\bse
\bearray
\begin{pmatrix} 0 & \chi \\ 0 & 0 \end{pmatrix} &=& \frac{1}{2 \sqrt{2}} \left ( 1+ \gamma n \right ) \gamma_5 , \\
\begin{pmatrix} 0 & 0 \\ \chi^\dagger & 0 \end{pmatrix} &=& \frac{1}{2 \sqrt{2}} \left ( 1 - \gamma n \right ) \gamma_5 .
\eearray
\ese
We also extract a factor of $m$ from each momentum vector and rename the resulting dimensionless variables with the same names as before.  The energy contribution takes the form
\bearray \label{deltaEa}
\Delta E_a &=& -4 i \pi^3 m \alpha^7 \int \frac{d^4 p}{(2 \pi)^4} \frac{d^4 q}{(2 \pi)^4} \frac{d^4 s}{(2 \pi)^4} \;
\mathrm{tr} \Bigl [ \gamma^\alpha \left ( \gamma(n-q)+1 \right ) \gamma^\beta \left ( 1-\gamma n \right ) \gamma_5 \Bigr ] \cr
&\times& \mathrm{tr} \Bigl [ \gamma^\mu \left ( \gamma(n-p)+1 \right ) \gamma^\nu \left ( 1+\gamma n \right ) \gamma_5 \Bigr ]
\times \mathrm{tr} \Bigl [ \gamma_\mu \left ( \gamma(s+p)+1 \right ) \gamma_\nu \left ( \gamma s+1 \right )  \cr
&\times& \gamma_\alpha \left ( \gamma (s+q)+1 \right )
\gamma_\beta \left ( \gamma(s+2n) +1 \right ) \Bigr ]
\times \Bigl [ p^2 (p-2n)^2 ((p-n)^2-1) ((p+s)^2-1) \cr
&\times&  (s^2-1) ((s+2n)^2-1) ((s+q)^2-1) q^2 (q-2n)^2 ((q-n)^2-1) \Bigr ]^{-1} .
\eearray
Using the properties of $\gamma_5$ in a trace, the first two traces can be expressed as
$\mathrm{tr} \bigl [ \gamma^\alpha \gamma q \gamma^\beta \gamma n \gamma_5 \bigr ]$ and
$-\mathrm{tr} \bigl [ \gamma^\mu \gamma p \gamma^\nu \gamma n \gamma_5 \bigr ]$
We choose to calculate the ground state energy shift ($n=1$) and will reinstate the full $n$ dependence at the end.  Expressions analogous to \eqref{deltaEa} can be obtained for each of the six diagrams that contribute to $\Delta E$.

Due to charge conjugation and other symmetries, only two of the six graphs in Fig.~\ref{fig1} are actually independent.  Charge conjugation can be used to show that the top three diagrams of Fig.~\ref{fig1} are equal to the bottom three, respectively.  The bottom three differ from the top diagrams only in the orientation of the light-by-light loop.  The identities can be obtained by using the fact that the trace of a transpose is the same as the trace of the original matrix and applying this identity to the light-by-light loop of, say, one of the diagrams on the bottom of Fig.~\ref{fig1}.  The transpose operation reverses the order of the matrices inside the trace.  Then factors of the charge conjugation matrix $C$ can be inserted using the identity $C C^{-1}=1$ and transposes can be exchanged for minus signs using $C^{-1} \gamma^{\mu T} C = -\gamma^\mu$.  Finally, a change of integration variable $s \rightarrow -s$ serves to confirm that Fig.~\ref{fig1}b=Fig.~\ref{fig1}a, Fig.~\ref{fig1}d=Fig.~\ref{fig1}c, and Fig.~\ref{fig1}f=Fig.~\ref{fig1}e.  The equality of Fig.~\ref{fig1}d with Fig.~\ref{fig1}a or of Fig.~\ref{fig1}b with Fig.~\ref{fig1}c can be shown by applying the substitutions $p \rightarrow 2n-p$, $s \rightarrow s+2n$, $\mu \leftrightarrow \nu$ and using the identity
\be
\mathrm{tr} \bigl [ \gamma^\mu \gamma p \gamma^\nu \gamma n \gamma_5 \bigr ] = \mathrm{tr} \bigl [ \gamma^\nu (-\gamma p) \gamma^\mu \gamma n \gamma_5 \bigr ].
\ee
In all, the first four graphs of Fig.~\ref{fig1} are equal to each other; we term these the ``planar'' graphs; and the last two, the ``crossed'' graphs, are also equal to each other.  So the complete energy correction can be written as $\Delta E = \Delta E_P + \Delta E_X$, where $\Delta E_P = 4 \Delta E_a$ is the planar contribution and $\Delta E_X = 2 \Delta E_e$ is the crossed contribution.

The particular challenge presented by this set of graphs is to separate the real from the imaginary parts of the energy correction.  These graphs have imaginary parts because the intermediate photons can be real photons, signaling the possibility of a physical decay process $\mathrm{p\!-\!Ps} \rightarrow 2 \gamma$ as corrected by light-by-light scattering of the outgoing photons.  In fact, the imaginary parts are relatively easy to compute, because they are simply related to the corresponding decay rate corrections, and have already been evaluated \cite{Czarnecki99,Adkins01}.  We first tried to perform the loop integrals using Feynman parameters, but were not able to separate the real and imaginary parts.  The technique we settled on was to do the energy integrals first using the residue theorem, after which we found it possible to identify and separate the real and imaginary parts, and then do the spatial integrals at the end.  This approach generated a number of terms, some of which contain denominators that vanish in the region of integration.  The singular behavior of individual terms is cancelled when all terms are included, but we found that quadruple precision was required in our routines for numerical integration over the spatial variables in order to make these cancellations manifest.


\section{The planar graphs}
\label{planar}

As a representative of the planar graphs we evaluated the diagram of Fig.~\ref{fig1}a (multiplied by four to account for all four planar graphs).  To do the energy integrals we used the residue theorem, so we needed to know whether the poles encountered were (slightly) above or (slightly) below the real energy axes.  For this purpose we made explicit the imaginary infinitesimals in the propagator factors that up until now had been implicit.  We found it advantageous to allow the infinitesimals to be different from one another.  The contribution of the planar graphs (four times the contribution of \eqref{deltaEa}) becomes
\be \label{deltaP}
\Delta E_P = \frac{2 m \alpha^7}{\pi^6} \int d^3 p \, d^3 q \, d^3 s \int \frac{d p_0}{2 \pi i} \frac{d q_0}{2 \pi i} \frac{d s_0}{2 \pi i} \; \frac{N_a}{D_a}
\ee
where
\bearray \label{numa}
N_a &=& \fourth \mathrm{tr} \Bigl [ \gamma^\alpha \gamma q\gamma^\beta \gamma n \gamma_5 \Bigr ]
\times \fourth \mathrm{tr} \Bigl [ \gamma^\mu \gamma p \gamma^\nu \gamma n \gamma_5 \Bigr ] \cr
&\times& \fourth \mathrm{tr} \Bigl [ \gamma_\mu \left ( \gamma(s+p)+1 \right ) \gamma_\nu \left ( \gamma s+1 \right ) \gamma_\alpha \left ( \gamma (s+q)+1 \right )
\gamma_\beta \left ( \gamma(s+2n) +1 \right ) \Bigr ]
\eearray
and
\bearray \label{dena}
D_a &=& (p^2+i \epsilon_1) ((p-2n)^2+i \epsilon_2) ((p-n)^2-1+i \epsilon_3) ((p+s)^2-1+i \epsilon_4) (s^2-1+i \epsilon_5) \cr
&\times& ((s+2n)^2-1+i \epsilon_6) ((s+q)^2-1+i \epsilon_7) (q^2+i \epsilon_8) ((q-2n)^2+i \epsilon_9) ((q-n)^2-1+i \epsilon_{10}) \cr
&=& (p_0-p+i \epsilon_1)(p_0+p-i \epsilon_1)(p_0-2-p+i \epsilon_2)(p_0-2+p-i \epsilon_2)(p_0-1-\omega_p+i \epsilon_3)(p_0-1+\omega_p-i \epsilon_3) \cr
&\times& (p_0+s_0-\omega_{ps}+i \epsilon_4)(p_0+s_0+\omega_{ps}-i \epsilon_4) (s_0-\omega_s+i \epsilon_5)(s_0+\omega_s-i \epsilon_5) \cr
&\times& (s_0+2-\omega_s+i \epsilon_6)(s_0+2+\omega_s-i \epsilon_6)(s_0+q_0-\omega_{sq}+i \epsilon_7)(s_0+q_0+\omega_{sq}-i \epsilon_7) \cr
&\times& (q_0-q+i \epsilon_8)(q_0+q-i \epsilon_8)(q_0-2-q+i \epsilon_9)(q_0-2+q-i \epsilon_9)(q_0-1-\omega_q+i \epsilon_{10})(q_0-1+\omega_q-i \epsilon_{10}) ,
\eearray
where $\omega_p=\sqrt{p^2+1}$, $\omega_s = \sqrt{s^2+1}$, $\omega_q = \sqrt{q^2+1}$, $\omega_{ps} = \sqrt{(\vec p + \vec s\,)^2+1}$, and $\omega_{sq} = \sqrt{(\vec s + \vec q\,)^2+1}$.  (From now through the end of this section the symbols $p$, $q$, and $s$ will refer to the magnitudes of the 3-vectors ($p \equiv \vert \vec p \, \vert$, $q \equiv \vert \vec q \, \vert$, $s \equiv \vert \vec s \, \vert$) except where explicitly stated otherwise.)
The traces were performed by the computer algebra system Reduce \cite{Hearn04}.  The poles integrals were done with the help of a routine written using Mathematica \cite{Wolfram12}.  Finally, the spatial momentum integrals were performed numerically using the adaptive Monte Carlo integration routine Vegas \cite{Lepage78}.  In setting up these last integrals, the direction of $\vec s$ was chosen to define the $z$-axis and the direction of $\vec p$ was then used to define the $xz$-plane.  The final integral was then six dimensional: three dimensions for the magnitudes of $s$, $p$, and $q$, one angle ($\theta_p$) for $\vec p$, and two angles ($\theta_q$ and $\phi_q$) for $\vec q$.  The energy contribution is independent of the three Euler angles describing the orientation in space of the triad ($\vec s,\, \vec p, \, \vec q \,$).  These Euler angles can be taken to be $\theta_s$ and $\phi_s$, defined with respect to an arbitrary external coordinate frame, and $\phi_p$, defined with an arbitrary origin.  The integral of the Euler angle volume element $d \phi_s \sin \theta_s d \theta_s \, d \phi_p$ contributes a factor $8 \pi^2$ to the energy.

The energy integrals were done using the residue theorem, one after the other.  A number of choices are required in order to perform the energy integrals.  The three integrals must be done in some order.  The contour containing the real axis must be closed with an infinite half-circle, which can be either in the upper or lower half-plane.  Finally, an ordering of the sizes of the infinitesimals has to be set.  An advantageous ordering can be used to minimize the number of terms appearing in the final result.  The result for the energy integral must not depend on which of the many possible choices is made, and this provides a strong check on our poles integration routine.  For example, when the $p_0$ integral is done first, closing in the lower half plane, then the $s_0$ integral, closing in the upper half plane, then the $q_0$ integral, closing in the lower half plane, and with the infinitesimals ordered so that $\epsilon_4$ is greater than any of $\epsilon_1$, $\epsilon_2$, $\epsilon_3$, and $\epsilon_7$ is greater than any of $\epsilon_1+\epsilon_4$, $\epsilon_2+\epsilon_4$, $\epsilon_3+\epsilon_4$, $\epsilon_5$, $\epsilon_6$, there are 56 terms of the form $N'/D'$ in the final result.

The next task was to separate the real and imaginary parts of the energy.  In terms of the standard three-loop hyperfine factor $m \alpha^7/\pi^3$ the energy takes the form
\be \Delta E_P = I_P \frac{m \alpha^7}{\pi^3} \ee
where
\be \label{integral_I_P}
I_P = \int ds \, d \theta_p \, d \theta_q \, d \phi_q \Bigl ( \frac{16}{\pi} s^2 \sin \theta_p \sin \theta_q \Bigr ) \int dp \, dq F(p,q)
\ee
with
\be
F(p,q) = \int \frac{d p_0}{2 \pi i} \frac{d q_0}{2 \pi i} \frac{d s_0}{2 \pi i} \frac{p^2 q^2 N_a}{D_a} .
\ee
All integration limits are standard: $0$ to $\infty$ for momentum magnitudes, $0$ to $\pi$ for polar angles, and $0$ to $2 \pi$ for azimuthal angles.  It was convenient to include all $p$ and $q$ dependence in $F(p,q)$, which depends implicitly on the variables $s$, $\theta_p$, $\theta_q$, $\phi_q$ as well as the ones explicitly shown.  There are a number of individual denominator factors in the $56$ terms that make up $F(p,q)$ that vanish within the region of integration.  Some, such as $2+\omega_s-p-\omega_{ps}-i (\epsilon_6-\epsilon_1-\epsilon_4)$, are innocuous even though there are values of $\vec p$ and $\vec s$ for which $2+\omega_s-p-\omega_{ps}=0$ because the singularities suggested by such a term cancel between the various contributions to $F(p,q)$.  For such terms the $\epsilon$'s can be set to zero and the cancellations will be manifest in the numerical evaluations (if done to sufficient precision).  The singularities coming from the factors $p-1-i (\epsilon_1+\epsilon_2)/2$ and $q-1-i (\epsilon_8+\epsilon_9)/2$, though, do not cancel.  These are exactly the singularities expected from the Cutkosky rules that occur when the intermediate photons are physical:
\be \label{reality_condition}
p_0>0, \; p^2=0; \; 2-p_0>0, \; (2n-p)^2=0,
\ee
(where in \eqref{reality_condition} $n$ and $p$ represent 4-vectors), which conditions imply $p=1$.  A similar condition holds for $q$.  We make these singularities visible by the expansion
\be
F(p,q) = \frac{A(p,q)}{(p-1-i \epsilon)(q-1-i \epsilon)} + \frac{B(p,q)}{(p-1-i \epsilon)} + \frac{C(p,q)}{(q-1-i \epsilon)} + D(p,q) ,
\ee
where it is necessary to know that $\epsilon$ is a positive infinitesimal but not its exact expression in terms of the $\epsilon_i$'s.  There is some ambiguity in the identification of the $A$--$D$ coefficients because, for example, the replacements $A(p,q) \rightarrow A(p,q)+(p-1)$, $C(p,q) \rightarrow C(p,q)-1$ leaves $F(p,q)$ unchanged.  This ambiguity does not affect the integrands that will eventually be obtained for the real and imaginary parts of $I_P$.  The actual separation was effected by the following procedure (used also in Ref. \cite{Adkins01}) that we illustrate for the integral of a function $f(x)$ of a single variable divided by $x-1-i \epsilon$:
\bearray
\int_0^\infty dx \, \frac{f(x)}{x-1-i \epsilon} &=& \int_0^2 dx \left ( \frac{f(1)}{x-1-i \epsilon} + \frac{f(x)-f(1)}{x-1-i \epsilon} \right ) + \int_2^\infty dx \frac{f(x)}{x-1-i \epsilon} \cr
&=& f(1) \int_0^2 \frac{dx}{x-1-i \epsilon} + \int_0^2 dx \frac{f(x)-f(1)}{x-1} + \int_2^\infty dx \frac{f(x)}{x-1} \cr
&=& i \pi f(1) + \int_0^\infty dx \frac{f(\tilde x)}{x-1} ,
\eearray
where
\be
f(\tilde x) = \begin{cases} f(x)-f(1) & \text{if} \; \; 0<x<2 \cr f(x) & \text{if} \; \; 2<x \end{cases} .
\ee
Application of this procedure to both the $p$ and $q$ integrals of \eqref{integral_I_P} leads to the form
\bearray \label{integral_I_PP}
I_P &=& \int ds \, d \theta_p \, d \theta_q \, d \phi_q \Bigl ( \frac{16}{\pi} s^2 \sin \theta_p \sin \theta_q \Bigr ) \cr
&\times& \Biggl [ \Bigl ( -\pi^2 A(1,1) + \int dp \, dq \left \{ \frac{A(\tilde p, \tilde q)}{(p-1)(q-1)} +\frac{B(\tilde p,q)}{(p-1)} + \frac{C(p, \tilde q)}{(q-1)} + D(p,q) \right \} \Bigr ) \cr
&+& i \pi \Bigl ( \int dp \left \{ \frac{A(\tilde p,1)}{(p-1)} + C(p,1) \right \} + \int dq \left \{ \frac{A(1, \tilde q)}{(q-1)} + B(1,q) \right \} \Bigr ) \Biggr ].
\eearray
The real part of the energy shift  is the sum of a four-dimensional integral
\be \label{I_P1}
I_{P1} \equiv \int ds \, d \theta_p \, d \theta_q \, d \phi_q \Bigl ( \frac{16}{\pi} s^2 \sin \theta_p \sin \theta_q \Bigr ) (-\pi^2) A(1,1) ,
\ee
and a six-dimensional integral
\bearray \label{I_P2}
I_{P2} &=& \int ds \, d \theta_p \, d \theta_q \, d \phi_q \, dp \, dq \, \Bigl ( \frac{16}{\pi} s^2 \sin \theta_p \sin \theta_q \Bigr ) \cr
&\times& \left \{ \frac{A(\tilde p, \tilde q)}{(p-1)(q-1)} +\frac{B(\tilde p,q)}{(p-1)} + \frac{C(p, \tilde q)}{(q-1)} + D(p,q) \right \}.
\eearray
The two five-dimensional integrals that comprise the imaginary part of $I_P$ are equal--as we have verified numerically--so that the full imaginary part can be written as
\be \label{I_P3}
I_{P3} = \int ds \, d \theta_p \, d \theta_q \, d \phi_q \, dp \, \Bigl ( \frac{16}{\pi} s^2 \sin \theta_p \sin \theta_q \Bigr ) (2 \pi) \left \{ \frac{A(\tilde p,1)}{(p-1)} + C(p,1) \right \} .
\ee

The cancellations among the various terms contributing to the integrals shown are so severe that even quadruple precision in Fortran is not sufficient to allow the intrinsic smoothness of the integrands to be apparent when a point chosen at random by the integration routine falls too close to one of the singularity surfaces that the individual terms possess.  Sufficient smoothness can be achieved though by averaging the integrands over the direction of $\vec q$ and that of $-\vec q$ \cite{Caswell79}.  The point is that
\be \int d^3 q \, f(\vec q \,) = \int d^3 q \, f(-\vec q \,) , \ee
so there is no reason not to replace an integrand $g(\theta_q, \phi_q)$ by $(g(\theta_q, \phi_q) + g(\pi-\theta_q, \pi+\phi_q))/2$.  This averaging leads to a smoother integrand that can be readily evaluated by standard numerical integration techniques.  The results for these integrals are given in Table~\ref{table1}.


\section{The crossed graphs}
\label{crossed}

We evaluated the crossed term by doubling the contribution of the graph of Fig.~\ref{fig1}e.  The formula for the energy $\Delta E_X$ is completely analogous to that of \eqref{deltaP} for $\Delta E_P$ except for being smaller by a factor of two (two graphs instead of four), having a different trace factor, and having a denominator $D_e$ that is identical to $D_a$ except for the replacement of $(s+2n)^2-1$ in $D_a$ by $(s+p+q-2n)^2-1$ in $D_e$.  The crossed contribution takes the form
\be \label{deltaX}
\Delta E_X = \frac{m \alpha^7}{\pi^6} \int d^3 p \, d^3 q \, d^3 s \int \frac{d p_0}{2 \pi i} \frac{d q_0}{2 \pi i} \frac{d s_0}{2 \pi i} \; \frac{N_e}{D_e}
\ee
with
\bearray \label{nume}
N_e &=& \fourth \mathrm{tr} \Bigl [ \gamma^\alpha \gamma q\gamma^\beta \gamma n \gamma_5 \Bigr ]
\times \fourth \mathrm{tr} \Bigl [ \gamma^\mu \gamma p \gamma^\nu \gamma n \gamma_5 \Bigr ] \cr
&\times& \fourth \mathrm{tr} \Bigl [ \gamma_\mu \left ( \gamma(s+p+q-2n)+1 \right ) \gamma_\beta \left ( \gamma (s+p)+1 \right ) \gamma_\nu \left ( \gamma s+1 \right )
\gamma_\alpha \left ( \gamma(s+q) +1 \right ) \Bigr ] .
\eearray
We performed the trace, the energy integrals, and the separation into real and imaginary parts as before.  The main difference between the evaluation of the contributions to $I_P$ given by \eqref{I_P1}, \eqref{I_P2}, and \eqref{I_P3} and the corresponding integrals $I_{X1}$, $I_{X2}$, and $I_{X3}$ for the crossed contribution is that the crossed integrands aren't so sensitive and do not require symmetrization in the directions of $\vec q$ in order to be successfully integrated.  Results for the crossed terms are given in Table~\ref{table1}.


\section{Results and discussion}
\label{results}

The results for all numerical integrals are shown in Table~\ref{table1}.  The four- and five-dimensional integrals were run with ten iterations of $10^8$ points after they were well adapted.  The six-dimensional integrals were more sensitive and problematic points occurred during iterations with $10^7$ points.  The decay rate of parapositronium into two photons with a light-by-light scattering correction is known, so the $\Delta \Gamma$ results can be compared against earlier values as a check of the present approach.  The old numbers were $1.988138(32)$ and $-0.694218(3)$ for the planar and crossed contributions to $\Delta \Gamma$ \cite{note1}.  The new results are consistent with the old ones, and the new decay rate correction
\be
\Delta \Gamma = 1.293945(19) \left ( \frac{\alpha}{\pi} \right )^2 \Gamma_0(\mathrm{p\!-\!Ps})
\ee
is twice as precise as the old value $1.29392(4)$.

\begin{table}[t]
\begin{center}
\caption{\label{table1} Results from numerical integration for the contributions to the energy shift due to light-by-light scattering in the two-photon-annihilation channel.  The planar contributions (from the diagrams of Fig.~\ref{fig1}a-d) and crossed contributions (from the diagrams of Fig.~\ref{fig1}e-f) are shown in separate columns.  All energy shifts are given in units of $m \alpha^7/\pi^3$.  Corrections to the decay rate $\Delta \Gamma=-2 \, \mathrm{Im}(\Delta E)$ are given in the last row in units of $(\alpha/\pi)^2 \Gamma_0(\mathrm{p\!-\!Ps})$ where $\Gamma_0(\mathrm{p\!-\!Ps}) = m \alpha^5/2$.}
\begin{ruledtabular}
\begin{tabular}{ccc}
Term & Planar & Crossed \\
\hline\noalign{\smallskip}
term 1 of  $\mathrm{Re}(\Delta E)$  &  2.054212(3) & -0.425630(1)   \\
term 2 of $\mathrm{Re}(\Delta E)$ &  -0.372170(67) & 0.327362(18) \\
total $\mathrm{Re}(\Delta E)$ & 1.682042(68) & -0.098268(19) \\
term 3 $= \mathrm{Im}(\Delta E)$,   &   -1.561502(14) &  0.545240(4) \\
$\Delta \Gamma$  & 1.988166(18) & -0.694221(5) \\
\end{tabular}
\end{ruledtabular}
\end{center}
\end{table}

The energy level correction due to light-by-light scattering in the two-photon-annihilation channel is
\be
\Delta E = 1.58377(8) \frac{m \alpha^7}{\pi^3} \frac{\delta_{\ell=0} \delta_{s=0}}{n^3}
\ee
where the $1/n^3$, orbital, and spin state dependence has been reinstated.  For the ground state this corrects the hfs (spin-triplet minus spin-singlet) by the amount
\be
\Delta E_{\rm{hfs}}= -6.95 kHz .
\ee
The numerical size of this contribution is small relative to present experimental precision, but it represents another essential step towards the goal of the complete calculation of all three-loop corrections to the positronium hfs.


\begin{acknowledgments}
We are grateful to Alexander Penin for a clarifying discussion on the ultrasoft contributions and to Calvin Stubbins for a useful suggestion on numerical integration.  We acknowledge the support of the National Science Foundation through Grant No. PHY-1404268 and of the Franklin \& Marshall College Grants Committee through the Hackman Scholars Program.
\end{acknowledgments}


     



\begin{thebibliography}{10}
\bibitem{note1} These numbers come from Table~I in \cite{Adkins01}.  The planar number in that reference was doubled to account for the four equal planar graphs instead of just two of them as was done in \cite{Adkins01}.
\end{thebibliography}
\end{document}